\documentclass[aps,twocolumn,superscriptaddress,showpacs]{revtex4}

\usepackage{graphicx,color}
\usepackage{dcolumn}
\usepackage{bm}
\usepackage{amsmath}
\usepackage{longtable}

\begin{document}

\preprint{}

\title{Lattice dynamics coupled to charge and spin degrees of freedom in the molecular dimer-Mott insulator $\kappa$-(BEDT-TTF)$_{2}$Cu[N(CN)$_{2}$]Cl}

\author{Masato~Matsuura$^{*}$}
  \affiliation{Neutron Science and Technology Center, Comprehensive Research Organization for Science and Society (CROSS), Tokai, Ibaraki 319-1106, Japan}
  \email{m_matsuura@cross.or.jp}
\author{Takahiko~Sasaki}
  \affiliation{Institute for Materials Research, Tohoku University, Sendai 980-8577, Japan}
\author{Satoshi~Iguchi}
  \affiliation{Institute for Materials Research, Tohoku University, Sendai 980-8577, Japan}
\author{Elena~Gati}
  \affiliation{Institute of Physics, SFB/TR49, Goethe-University Frankfurt, 60438 Frankfurt (M), Germany}
\author{Jens M\"uller}
  \affiliation{Institute of Physics, SFB/TR49, Goethe-University Frankfurt, 60438 Frankfurt (M), Germany}
\author{Oliver Stockert}
  \affiliation{Max-Planck-Institut f\"ur Chemische Physik fester Stoffe, D-01187 Dresden, Germany}
\author{Andrea Piovano}
  \affiliation{Institut Laue-Langevin, 6 rue Jules Horowitz, 38042 Grenoble Cedex 9, France}
\author{Martin B\"ohm}
  \affiliation{Institut Laue-Langevin, 6 rue Jules Horowitz, 38042 Grenoble Cedex 9, France}
\author{Jitae T. Park}
  \affiliation{Heinz Maier-Leibnitz Zentrum (MLZ), Technische Universit\"at M\"unchen, Lichtenbergstr. 1, 85748 Garching, Germany}
\author{Sananda Biswas}
  \affiliation{Institute for Theoretical Physics, SFB/TR49, Goethe-University Frankfurt, 60438 Frankfurt (M), Germany}
\author{Stephen M. Winter}
  \affiliation{Institute for Theoretical Physics, SFB/TR49, Goethe-University Frankfurt, 60438 Frankfurt (M), Germany}
\author{Roser Valent\'i}
  \affiliation{Institute for Theoretical Physics, SFB/TR49, Goethe-University Frankfurt, 60438 Frankfurt (M), Germany}
\author{Akiko~Nakao}
  \affiliation{Neutron Science and Technology Center, Comprehensive Research Organization for Science and Society (CROSS), Tokai, Ibaraki 319-1106, Japan}
\author{Michael Lang}
  \affiliation{Institute of Physics, SFB/TR49, Goethe-University Frankfurt, 60438 Frankfurt (M), Germany}

\begin{abstract}
Inelastic neutron scattering measurements on the molecular dimer-Mott insulator $\kappa$-(BEDT-TTF)$_{2}$Cu[N(CN)$_{2}$]Cl
reveal a phonon anomaly in a wide temperature range. Starting from $T_{\rm ins}\sim50$-$60$~K where the charge gap opens, 
the low-lying optical phonon modes become overdamped upon cooling towards the antiferromagnetic ordering temperature $T_\mathrm{N} = 27$\,K, 
where also a ferroelectric ordering at $T_{\rm FE} \approx T_{\rm N}$ occurs.
Conversely, the phonon damping becomes small again when spins and charges are ordered below $T_\mathrm{N}$,
while no change of the lattice symmetry is observed across $T_\mathrm{N}$ in neutron diffraction measurements.
We assign the phonon anomalies to structural fluctuations coupled to charge and spin degrees of freedom in the BEDT-TTF molecules.
\end{abstract}


\maketitle

Electronic ferroelectricity, where electrons and their interactions play the key role,
has been in the focus of recent scientific efforts~\cite{Brink08, Horiuchi08, Ishihara11}.
Whereas conventional ferroelectricity is driven by shifts in the atomic positions,
electronic ferroelectricity originates from electronic degrees of freedom, such as spin and charge,
which offers an alternative route to control the system's ferroelectric properties.
Electronic ferroelectricity, driven by spin degrees of freedom,
is often found in frustrated magnetic systems with non-collinear spin structures
~\cite{Ishihara11, Kimura03}.
In case of the charge-driven variant, the electric dipoles arise from
charge order or charge disproportionation in combination with dimerization,
which has been found in inorganic oxides~\cite{Ikeda05}
and organic charge-transfer salts~\cite{Chow2000,Yamamoto08,Abdel10,Iguchi13,Gati2018}. 

The $\kappa$-(BEDT-TTF)$_{2}X$ family,
where BEDT-TTF is bis-(ethylenedithio)tetrathiafulvalene C$_{6}$S$_{8}$[(CH$_{2}$)$_{2}$]$_{2}$,
is known to comprise bandwidth-controlled dimer-Mott systems where pairs of strongly interacting BEDT-TTF (in short ET) molecules form the dimers.
In the dimer-Mott insulator picture, one $\pi$-hole carrier with spin $S=1/2$ is localized on a molecular dimer unit. 
The charge degrees of freedom may become active when this localization is no longer symmetric 
with respect to the center of the dimer but rather adopts an asymmetric state characterized by a charge disproportionation within the dimer~\cite{Naka10, Hotta12}.
Such a charge disproportionation scenario was suggested as the origin of the relaxor-type dielectric anomaly
observed in the quantum-spin-liquid-candidate material $X$ = Cu$_{2}$(CN)$_{3}$ ($\kappa$-CN)~\cite{Shimizu03, Abdel10} -- a suggestion which
has created enormous attention as it highlights the important role of the intra-dimer charge degrees of freedom.
Since relaxor ferroelectrics are known to consist of nanometer-sized domains,
the relaxor-like dielectric anomaly in $\kappa$-CN suggests the presence of an inhomogeneous charge disproportionation.
Recently, Lunkenheimer {\it et al.} have reported clear ferroelectric signatures in the related dimer-Mott system $X$ = Cu[N(CN)$_{2}$]Cl ($\kappa$-Cl)
around the antiferromagnetic ordering temperature $T_\mathrm{N} = 27$\,K~\cite{Lunkenheimer12,Lang2014}. 
As in this system long-range ferroelectricity of order-disorder type is observed at $T_{\rm FE} \approx T_{\rm N}$, 
$\kappa$-Cl represents an ideal system to study the coupling of the charge- to the spin- and lattice degrees of freedom in a dimer-Mott insulator.

It is fair to say that the origin of the electric dipoles in these dimer-Mott insulators is still under debate. 
Whereas for $\kappa$-Cl and $\kappa$-CN a definite proof of charge disproportionation is still missing~\cite{Sedlmeier12}, 
clear evidence for charge order within the ET dimers has recently been found 
for the more weakly dimerized compound $X$ = Hg(SCN)$_{2}$Cl \cite{Drichko2014}, 
making this system a prime candidate for electronically-driven ferroelectricity within the $\kappa$-(ET)$_{2}X$ family \cite{Gati2018}.

Given that there is a finite electron-lattice coupling, fluctuations of the electric dipoles are expected to give rise to anomalies in the lattice dynamics 
which can be sensitively probed by neutron scattering. In fact, for relaxor ferroelectrics, 
neutron scattering studies have been able to reveal a phonon anomaly upon
the appearance of inhomogeneous and fluctuating polar domains~\cite{Gehring00}.

Unfortunately, systematic inelastic neutron scattering (INS) studies
on organic charge-transfer salts have often been hampered due to the lack of sufficiently large single crystals with only a few exceptions:
for instance, a sizable phonon renormalization effect on entering the superconducting state was reported
for the organic superconductor $\kappa$-(ET)$_{2}$Cu(NCS)$_{2}$~\cite{Pintschovius97}.
Thanks to major recent improvements in focusing the neutron beam, however, the situation has improved considerably. 
As we demonstrate in this work, the largely enhanced neutron flux at the sample position in state-of-the-art triple-axis-spectrometers~\cite{IN8, PUMA} 
now enables such INS studies to be performed even on small single crystals of organic charge-transfer salts.
Here we report an INS study of the lattice dynamics and its coupling to the charge- and spin degrees of freedom for the dimer-Mott insulator $\kappa$-Cl. 
By using an array of co-aligned single crystals of deuterated $\kappa$-Cl with total mass of 7~mg and 9~mg, 
we were able to detect clear phonon signals the amplitude of which shows a striking variation upon changing the temperature. 
We found that the low-lying optical phonon modes at 2.6~meV become damped
below the onset temperature of the dimer-Mott insulating state in which the $\pi$-carriers start to localize on the dimer sites.
This phonon damping becomes small again on cooling below $T_\mathrm{N}$.
In contrast to conventional displacive ferroelectrics, however, there is no divergence of the phonon intensity
nor any change in the lattice symmetry at $T_\mathrm{N}$ where also the dielectric anomaly was observed.
Thus, in $\kappa$-Cl the lattice is clearly coupled to the charge- and spin degrees of freedom 
but appears not to be the driving force of the antiferromagnetic/ferroelectric phase transition at $T_\mathrm{N}/T_{\rm FE}$.

Deuterated single crystals of $\kappa$-(ET)$_2$Cu[N(CN)$_2$]Cl were grown by
electrochemical crystallization.
The N\'eel temperature was determined to be $T_\mathrm{N} = 27$\,K from magnetic susceptibility measurements as shown in Fig.~\ref{fig3}(d),
which is identical to the ordering temperature reported for hydrogenated $\kappa$-Cl~\cite{Miyagawa95}.
INS experiments were performed on the
triple-axis spectrometers  IN8 at the Institut Laue-Langevin~\cite{IN8}
and PUMA at the Heinz Maier-Leibnitz Zentrum~\cite{PUMA}.
All data were collected with a fixed final neutron energy of 14.7~meV
using a doubly focused Cu analyzer for IN8 and
a doubly focused pyrolytic graphite (PG) analyzer for PUMA.
The initial neutron energy was selected by
a doubly focused PG monochromator for both IN8 and PUMA.
A PG filter was placed in front of the analyzer
to suppress the scattering of higher-order neutrons.
To improve the signal-to-noise ratio,
two single crystals with a total mass of 7~mg,
and six single crystals with a total mass of 9~mg
were co-aligned for the neutron experiments on IN8 and PUMA, respectively.
In all experiments, the samples were slowly cooled with 1~K/min around $T=75$~K
to minimize disorder in the ET molecules' ethylene endgroup orientations~\cite{JMueller2002,JMueller2015}.
The single crystals were mounted
so as to access the ($h0l$) and ($hk0$) scattering plane
for INS and neutron diffraction measurements, respectively.
Throughout this paper, we label the
momentum transfer  in units of the reciprocal lattice vectors
$a^{*} = 0.484$~\AA$^{-1}$, $b^{*} = 0.210$~\AA$^{-1}$ and $c^{*}=0.741$~\AA$^{-1}$.
The instrumental energy resolution for IN8 linearly increases 
from 0.45~meV ($E=0$) to 0.84~meV ($E=10$~meV). 
To relate the fit parameters to the scattering function of the sample, 
convolution of the instrumental resolution at $E=3$~meV has been included and computed 
using the RESTRAX simulation package~\cite{RESTRAX}.
We assume flat dispersions within $Q$-width of the resolution ellipsoid 
since optical dispersions close to the $\Gamma$-point are flat.

\begin{figure}
\includegraphics[keepaspectratio=true,width=7cm,clip]{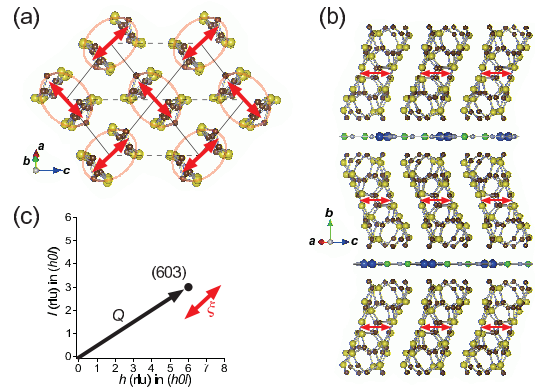}
\caption{\label{fig1} (color online)
Crystal structure of  $\kappa$-(ET)$_{2}$Cu[N(CN)$_{2}$]Cl:
(a) Top view of the ET layer and (b) side view of the layered structure.
(c) Wave vector $\mbox{\boldmath $Q$}$=(603) chosen for phonon measurements in the ($h0l$) scattering plane.
Ellipses in (a) indicate ET dimers.
Thick arrows show schematically the breathing mode with polarization vector $\mbox{\boldmath $\xi$}$
of ET molecules mainly detected at (603).
}
\end{figure}

Figures~\ref{fig1} (a) and (b) show the crystal structure of $\kappa$-Cl
consisting of layers of ET molecules separated by thin anion sheets. The
ET molecules form dimers, resulting in a dimer-Mott insulating ground state.
Since the distance between the ET molecules within the dimer reflects the degree of dimerization,
some of the modes are expected to couple more strongly to the electronic degrees of freedom than others. 
One of such modes is a breathing mode of the ET dimers, 
shown schematically by thick arrows in Figs. 1(a) and (b).
The scattering intensity of phonons in neutron scattering 
is proportional to ($\mbox{\boldmath $Q$}\cdot \mbox{\boldmath $\xi$}$)$^{2}$,
where $\mbox{\boldmath $Q$}$ is the momentum transfers between the initial and final state of the neutron,
and  $\xi$ is the polarization vector of the phonon mode.
Thus, the breathing of the ET dimers can best be measured
when $\mbox{\boldmath $Q$}$ is large and parallel to $\xi$ of the breathing mode
as shown in Fig.~\ref{fig1}(c).
We measured the phonon spectra mainly at (603) to detect changes in the
low-lying vibrational modes which are likely coupled to the charge- and spin degrees of freedom;
changes will be detected for any vibrational mode which has a component parallel to [603].

\begin{figure}
\includegraphics[keepaspectratio=true,width=6cm,clip]{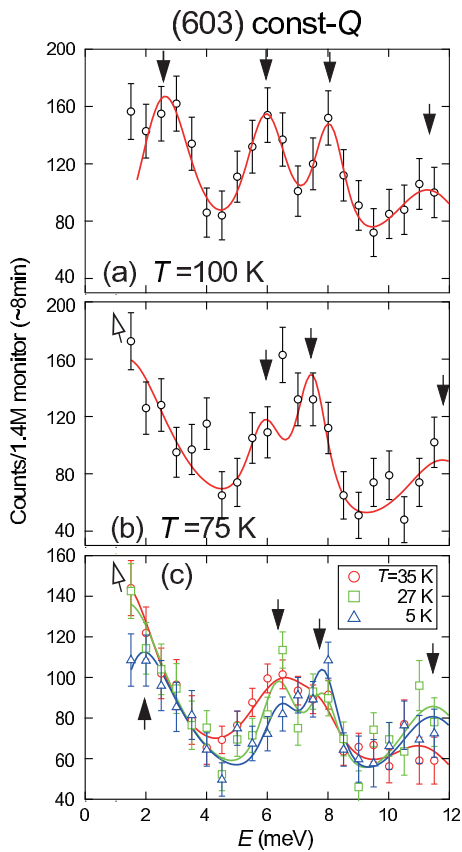}
\caption{\label{fig2} (color online)
Temperature dependences of constant-Q scans at (603) measured by using IN8.
The solid lines are fits to four damped harmonic oscillator functions
at $E\sim2.6$, 6, 8, and 11~meV.
The instrumental resolution was convolved to the model cross sections
assuming flat dispersion within the resolution.
Closed arrows indicate underdamped phonon modes, while open arrows show the
overdamped phonon mode for $T_\mathrm{N}<T\leq75$~K.
}
\end{figure}
Figures~\ref{fig2}(a)-(c) show constant-$Q$ scans at (603)
measured at various temperatures.
At $T=100$~K,  we observe clear phonon peaks at $E=2.6$, 6, 8, and 11~meV
shown by the closed arrows.
(See Figs.~S1 and S2 in the Supplemental Material for the details of the dispersion~\cite{Supplement}).
The peak width for the low-lying modes at $E=2.6$~meV (full-width-at-half-maximum; 2.3~meV)
is considerably larger than the energy resolution (0.5~meV) indicating a finite lifetime
due to phonon-phonon or electron-phonon interactions.
For an anharmonic phonon, the energy dependence of the scattering function can be expressed by
by the damped harmonic-oscillator function:~\cite{Gesi72}
\begin{equation}
\frac{\Gamma_{q}\hbar\omega}{[\hbar^{2} (\omega^{2}-\omega^{2}_{q})]^{2}+(\Gamma_{q}\hbar\omega)^{2}}
\label{eq1}
\end{equation}
where $\hbar\omega$ is the energy transfers between the initial and final state of the neutron, 
and $\Gamma_{q}$ denotes the damping factor.
The lifetime of the phonons is inversely proportional to $\Gamma_{q}$.
The enhanced phonon width, when compared to the energy resolution, suggests a strongly anharmonic lattice, 
consistent with the observation of large expansion coefficients
in the $\kappa$-(ET)$_{2}X$ family~\cite{JMueller2002}.
On cooling, the well-resolved peak from the low-lying optical modes at 2.6~meV changes into
a sloped signal at $T=75$~K, whereas the three modes at 6, 8, and 11~meV
remain at almost the same energy [Fig.~\ref{fig2}(b)].
When the damping factor $\Gamma_{q}$ becomes comparable to $\omega_{q}$,
the phonon spectrum changes into a single peak at $E=0$.
The sloped spectrum can be explained by an increasing $\Gamma_{q}$
for the low-lying optical modes, which indicates the short lifetime of these modes.
Similar phonon spectra at low energies
are observed down to $T_{\rm N} = 27$\,K [Fig.~\ref{fig2}(c)].
Below $T_{\rm N}$, the truly inelastic nature of the low-lying optical modes become again visible.

\begin{figure}
\includegraphics[keepaspectratio=true,width=7.5cm,clip]{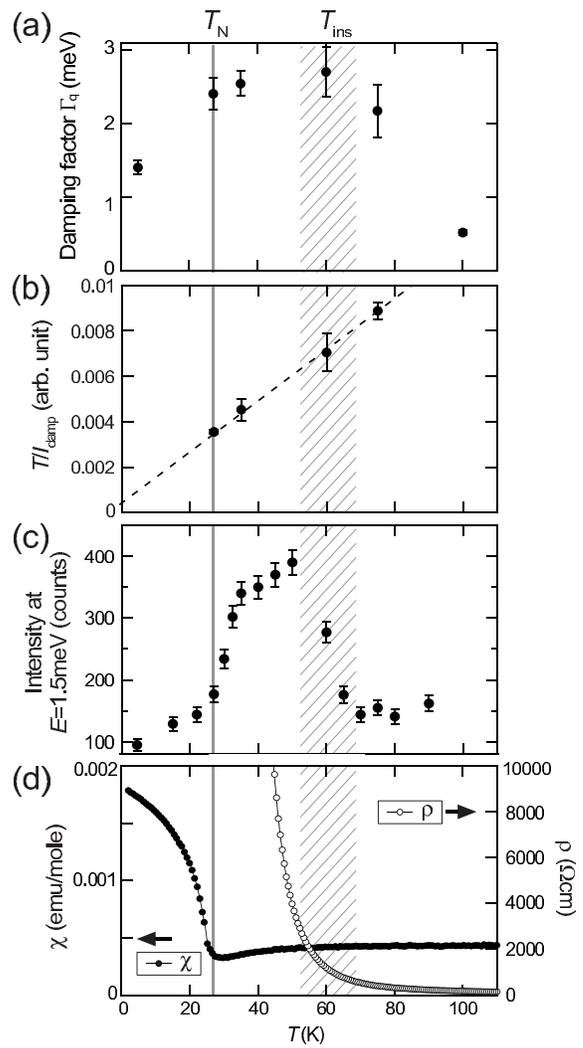}
\caption{\label{fig3} (color online)
Temperature dependence of (a) the damping factor $\Gamma_{q}$ for the low-lying optical modes at (603),
(b) $T$/$I_\mathrm{damp}$ for the low-lying optical modes (see text for details),
(c) the intensity at $E=1.5$~meV at (603), (d) the out-of-plane electrical resistivity ($\parallel b$), and out-of-plane
dc magnetic susceptibility ($\parallel b$) at $\mu_{0}H=0.5$~T.
The hatched area at $\sim50$-$60$~K represent the cross-over temperature $T_{\rm ins}$ as described in the text.
The broken line in (b) is a linear fit.
Data in (a) were measured at IN8 whilst data in (c) were obtained at PUMA.
}
\end{figure}

Figure~\ref{fig3}(a) shows the temperature dependence of the damping factor $\Gamma_{q}$
for the low-lying optical modes.
Note that $\Gamma_{q}$ is comparable to $\omega_{q}$ in a wide temperature range
for $T_\mathrm{N}<T<75$~K, indicating the damped nature of the low-lying optical modes.
In addition to the phonon anomaly,
we find that the energy width of the Bragg peaks is broadened
in the same temperature range (Fig.~S3 in the Supplemental Material),
suggesting an unstable lattice in this temperature region.
Thus, the lattice of $\kappa$-Cl shows anomalous behavior
in a wide temperature range above $T_\mathrm{N}$.

Overdamped soft modes are often seen near structural phase transitions in a variety of materials.
In the overdamped regime, the scattering function
$S(\mbox{\boldmath $Q$},\omega)$ is approximated as~\cite{Shirane}:
\begin{equation}
S(\mbox{\boldmath $Q$},\omega)=\frac{(2\pi)^3}{v_{0}} |F(\mbox{\boldmath $Q$})|^{2}\frac{k_{B}T}{\hbar\omega^{2}_{q}} \frac{1}{\pi} \frac{\gamma_{q}}{\omega^2+\gamma^{2}_{q}},
\label{eq2}
\end{equation}
where $v_{0}$ is the unit cell volume, $F(\mbox{\boldmath $Q$})$ is a dynamical structure factor, 
and $\gamma_{q}=\frac{\omega^{2}_{q}}{2\Gamma_{q}}$.
Thus, the integrated intensity of an overdamped soft mode ($I_\mathrm{damp}=\int S(\mbox{\boldmath $Q$},\omega) d\omega$)
is proportional to $k_{B}T/\omega^{2}_{q}$ .
As $\omega_{q=0}$ goes to zero at the transition temperature,
the intensity of the damped soft mode diverges and $T$/$I_\mathrm{damp}$ goes to zero.
Figure~\ref{fig3}(b) shows the temperature dependence of $T$/$I_\mathrm{damp}$
for the low-lying optical modes.
Clearly, $T$/$I_\mathrm{damp}$ does not vanish at $T_\mathrm{N}$,
which indicates that the structural change is not the primary order parameter
for the phase transition at $T_\mathrm{N}$.
This is consistent with the observation of only a small anomaly in the thermal expansion at $T_\mathrm{N} \approx T_{\rm FE}$
for $\kappa$-Cl~\cite{JMueller2002}.
Instead, other degrees of freedom show divergent behavior at $T_\mathrm{N}$, namely
($T_{1}T$)$^{-1}$ of $^{1}H$ NMR for the spin~\cite{Miyagawa95}
and the dielectric constant for the charge degrees of freedom~\cite{Lunkenheimer12}.

Note that the overdamped phonon appears already at $60 - 75$~K,
which is much higher than $T_\mathrm{N}$.
Figure~\ref{fig3}(c) shows the temperature dependence of the scattering intensity at
$Q=$(603) and $E=1.5$~meV.
The effect of the phonon anomaly is clearly seen as an enhancement of
the intensity at low energies ($E=1.5$~meV)
for $T_\mathrm{N}<T\lesssim60$~K.
The coupling between a phonon mode and
a relaxation mode of different origin (represented by a pseudospin) was discussed 
by Yamada within a pseudospin-phonon coupling model~\cite{Yamada74}.
In this model,
the relaxation mode and the phonon mode become strongly coupled
when the characteristic frequency and wave vector
of the pseudospin's relaxation mode roughly match those of the phonon mode,
which results in a broadening of the phonon in energy.		
The structural fluctuations between the two ethylene endgroup orientations
of the ET molecules, known to occur in the $\kappa$-phase (ET)$_{2}X$ salts~\cite{JMueller2015,Urs91}, 
however cannot be the origin of the observed phonon anomaly
since the ethylene endgroup motion freezes out in a glassy fashion below about 75~K.

On the other hand, the onset temperature of the phonon anomaly roughly
coincides with the rapid increase in
the electrical resistivity below $T_{\rm ins}\sim50$-$60$~K, as shown in Fig.~\ref{fig3}(d).
Optical conductivity measurements have revealed that
the charge gap starts to open below $T_{\rm ins}$~\cite{Kornelsen92, Sasaki04},
indicating charge localization at the ET dimer site.
Even after the itinerant $\pi$-carriers become localized at the dimer site below $T_{\rm ins}$,
the distribution of charge within the dimer remains as an active degree of freedom.
The coincidence of the onset temperature of the phonon anomaly and $T_{\rm ins}$
suggests a coupling between the lattice and this intra-dimer charge degree of freedom.
A similar scenario was observed in the organic superconductor
$\kappa$-(ET)$_{2}$Cu(NCS)$_{2}$ for the lowest optical mode~\cite{Pintschovius97}.
We thus expect that the low-lying optical modes are the key modes
where the electron-phonon coupling becomes manifest in these organic salts containing ET molecules.

In an attempt to identify the low-lying optical modes with the theoretically obtained vibrational frequencies, 
we considered the phonon frequencies of the closely related system $\kappa$-CN~\cite{Dressel16} 
which has a similar arrangement of anion and ET layers as $\kappa$-Cl, 
but exhibits considerably less phonon modes.
The calculated lowest optical mode in $\kappa$-CN at $q=0$ 
has an energy of 3~meV and two of the ET-dimer breathing modes, 
involving also the movements of the anion layers, lie at energies of 4.1 meV and 4.7 meV. 
The $\kappa$-Cl system differs from $\kappa$-CN mainly in two ways: 
(i) the ET-dimers in $\kappa$-Cl are alternately stacked along the $b$ axis, 
thus doubling the unit cell along that direction and 
(ii) the anion layers consist of chains arranged in a polymeric zigzag pattern, 
instead of having a more rigid two-dimensional network in $\kappa$-CN. 
It is to be expected that these factors will lead to significantly softer spring-constants 
in $\kappa$-Cl compared to $\kappa$-CN.
Such a rescaling of the phonon modes would affect all the phonons in the low-frequency regime.
The lowest mode in $\kappa$-CN may shift toward very low energies
and merge with the elastic peak,
which may allow to identify the observed low-lying optical modes 
in $\kappa$-Cl as an ET-dimer breathing mode.

Below $T_\mathrm{N}$,
the damping of the low-lying optical modes become small
as shown in Fig.~\ref{fig2}(c) and Fig.~\ref{fig3}(a).
According to the pseudospin-coupling model~\cite{Yamada74},
the reduction in the damping factor $\Gamma_{q}$ is due to the
decoupling between the lattice and the pseudospins as a consequence of a critical slowing down of the pseudospin fluctuations.
Since both dielectric and antiferromagnetic fluctuations freeze out below $T_\mathrm{N}$~\cite{Lunkenheimer12},
the recovery of a truly inelastic phonon peak
suggests a close correlation between lattice-, spin- and charge degrees of freedom
at the phase transition at $T_\mathrm{N} \approx T_{\rm FE}$.

In order to check for the possibility of a change in the crystallographic symmetry at $T_\mathrm{N}$, 
we performed detailed neutron diffraction measurements on a deuterated single crystal of $\kappa$-Cl
at $T=35$~K ($>T_\mathrm{N}$) and 4~K ($<T_\mathrm{N}$) (Fig.~S4 in the Supplemental Material). 
As explained in detail in the supplemental material, 
we did not find any indications for a crystallographic symmetry lowering at $T_\mathrm{N} \approx T_{\rm FE}$. 
This supports the picture of electronic ferroelectricity, 
where instead of the lattice, the spin- or charge degrees of freedom are the driving force of the phase transition in $\kappa$-Cl. 

In discussing our results, 
we recall that recent vibrational spectroscopy studies failed to detect clear signatures 
of a charge disproportionation in $\kappa$-Cl and $\kappa$-CN~\cite{Sedlmeier12}.
On the other hand, our finding of overdamped modes for $T_\mathrm{N}<T\lesssim T_{\rm ins}$
strongly suggests a close coupling between the lattice and the intra-dimer charge degrees of freedom.
According to the pseudospin-coupling model~\cite{Yamada74},
the characteristic energy of the charge fluctuations is expected to be
in the same range as the low-lying optical modes, $1 - 2$~meV,
for $T_\mathrm{N}<T\lesssim T_{\rm ins}$.
This energy scale is two orders of magnitude smaller than that
of the charge-sensitive mode studied in the above-mentioned vibrational spectroscopy experiments.
Furthermore, a finite DC conductivity is observed even in the "Mott-insulating" state below $\sim T_{\rm ins}$
reflecting some degree of remaining itinerancy of the fluctuating $\pi$-electrons.
The complex and seemingly contradictory picture of the low energy charge dynamics reported so far 
is likely due to the different time-scales of the investigated characteristic mode.

To summarize, by studying the spectra of selected phonons of the dimer-Mott insulator 
$\kappa$-(ET)$_2$Cu[N(CN)$_2$]Cl as a function of temperature, 
we found clear renormalization effects which can be associated with charge fluctuations. 
We argue that the overdamped optical phonon modes, observed in a wide temperature
 range from $T_{\rm ins}\sim50$-$60$~K down to $T_\mathrm{N}  \approx  T_{\rm FE}$, 
result from a coupling of the lattice to the intra-dimer charge degrees of freedom. 
We consider these inelastic neutron scattering results as an important step 
which may trigger further systematic studies on the lattice dynamics and its coupling to 
the electronic degrees of freedom in the family of organic charge-transfer salts.

We are grateful to M. Naka and S. Ishihara for helpful discussions.
We also thank O. Sobolev, M. Kurosu, R. Kobayashi, B. Hartmann, T. Ohhara, and K. Munakata
for their help in the experiments.
The neutron experiments were performed with the approval of ILL (7-01-401),
MLZ (11879), and J-PARC MLF (2017B0201).
The crystal structures in Fig.~\ref{fig1} are produced by VESTA software\cite{Momma_11}.
This study was financially supported by
Grants-in-Aid for Scientific Research 
(Grants Nos. 25287080, 19H01833, 16K05430, and 18H04298)
from the Japan Society for the Promotion of Science.

\end{document}



\title{Supplemental material\\ for\\ Lattice dynamics coupled with charge and spin degrees of freedom in the molecular dimer Mott insulator $\kappa$-(BEDT-TTF)$_{2}$Cu[N(CN)$_{2}$]Cl}

\author{Masato~Matsuura$^{*}$}
  \affiliation{Research Center for Neutron Science and Technology, Comprehensive Research Organization for Science and Society (CROSS), Tokai, Ibaraki 319-1106, Japan}
  \email{m_matsuura@cross.or.jp}
\author{Takahiko~Sasaki}
  \affiliation{Institute for Materials Research, Tohoku University, Sendai 980-8577, Japan}
\author{Satoshi~Iguchi}
  \affiliation{Institute for Materials Research, Tohoku University, Sendai 980-8577, Japan}
\author{Elena~Gati}
  \affiliation{Institute of Physics, SFB/TR49, Goethe-University Frankfurt, 60438 Frankfurt (M), Germany}
\author{Jens M\"uller}
  \affiliation{Institute of Physics, SFB/TR49, Goethe-University Frankfurt, 60438 Frankfurt (M), Germany}
\author{Oliver Stockert}
  \affiliation{Max-Planck-Institut f\"ur Chemische Physik fester Stoffe, D-01187 Dresden, Germany}
\author{Andrea Piovano}
  \affiliation{Institut Laue-Langevin, 6 rue Jules Horowitz, 38042 Grenoble Cedex 9, France}
\author{Martin B\"ohm}
  \affiliation{Institut Laue-Langevin, 6 rue Jules Horowitz, 38042 Grenoble Cedex 9, France}
\author{Jitae T. Park}
  \affiliation{Heinz Maier-Leibnitz Zentrum (MLZ), Technische Universit\"at M\"unchen, Lichtenbergstr. 1, 85748 Garching, Germany}
\author{Sananda Biswas}
  \affiliation{Institute for Theoretical Physics, SFB/TR49, Goethe-University Frankfurt, 60438 Frankfurt (M), Germany}
\author{Stephen M. Winter}
  \affiliation{Institute for Theoretical Physics, SFB/TR49, Goethe-University Frankfurt, 60438 Frankfurt (M), Germany}
\author{Roser Valent\'i}
  \affiliation{Institute for Theoretical Physics, SFB/TR49, Goethe-University Frankfurt, 60438 Frankfurt (M), Germany}
\author{Akiko~Nakao}
  \affiliation{Research Center for Neutron Science and Technology, Comprehensive Research Organization for Science and Society (CROSS), Tokai, Ibaraki 319-1106, Japan}
\author{Michael Lang}
  \affiliation{Institute of Physics, SFB/TR49, Goethe-University Frankfurt, 60438 Frankfurt (M), Germany}
\date{\today}

\maketitle


\section{Supplementary Material}

\begin{figure}
\includegraphics[keepaspectratio=true,width=8.0cm,clip]{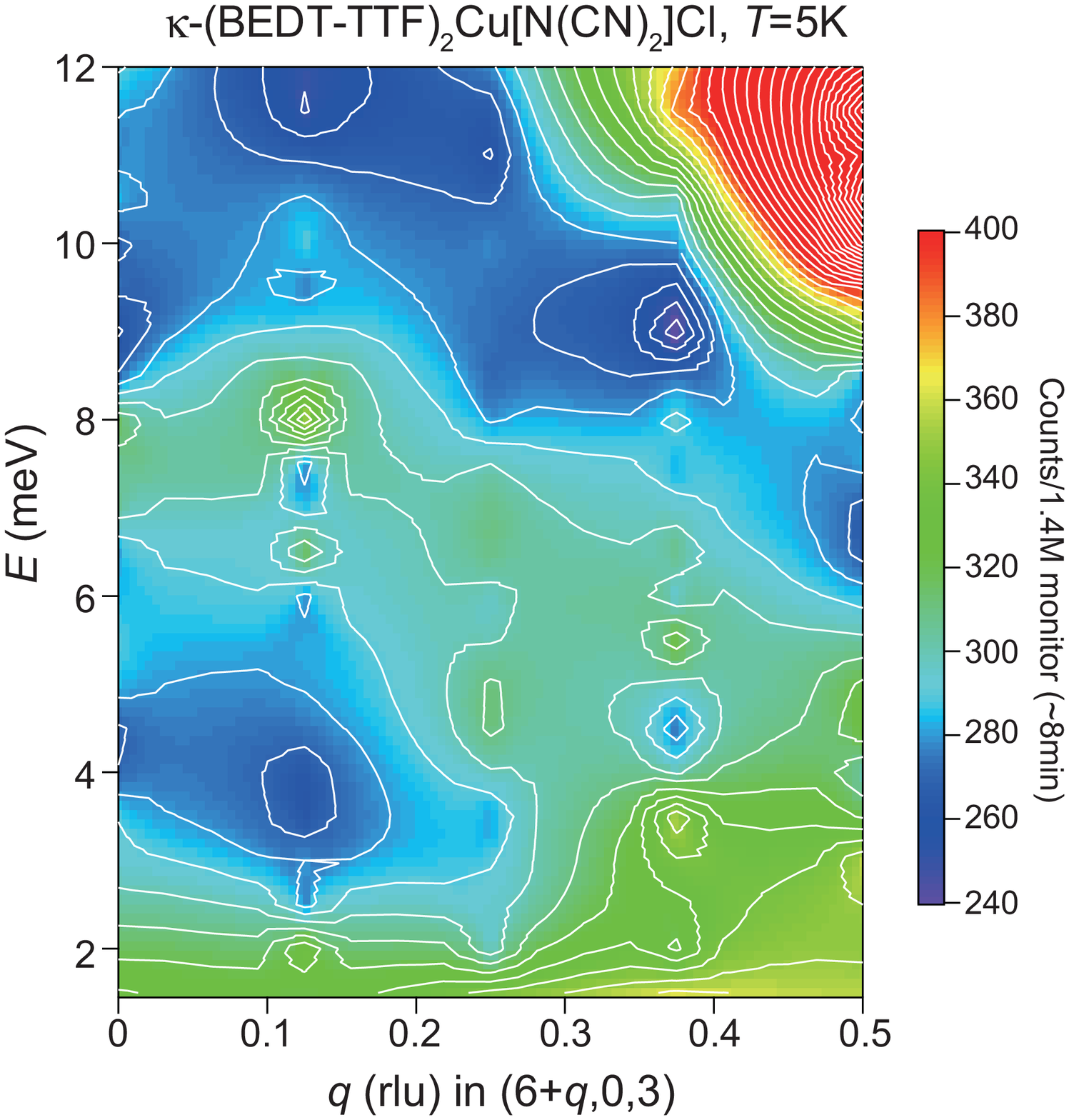}
\caption{\label{fig1}
Contour map of phonon scattering intensity near (603) along the [100] direction
measured at $T=5$~K.
The contour map is reproduced from constant-Q scans 
shown in Fig.~\ref{fig2}. 
}
\end{figure}

\begin{figure*}
\includegraphics[keepaspectratio=true,width=14.0cm,clip]{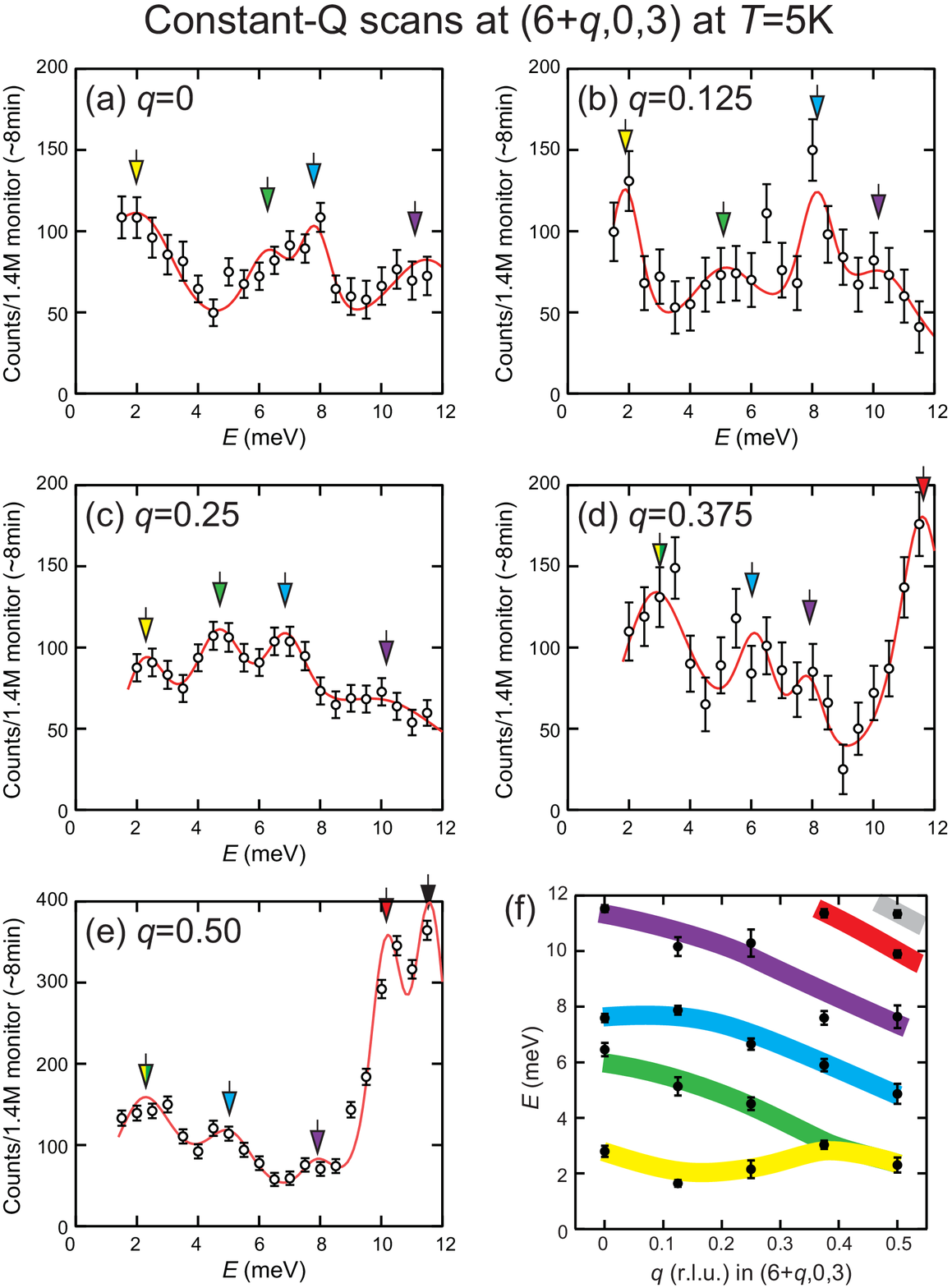}
\caption{\label{fig2} 
(a)-(e) Constant-Q cuts of $\kappa$-Cl at ($6+q$,0,3) measured at $T=5$~K.
The solid lines are fits to damped harmonic oscillator functions.
We convoluted the instrumental resolution to the model cross sections
assuming flat dispersions by using the RESTRAX simulation package~\cite{RESTRAX}.
Arrows indicate energies of the phonon modes.
Data were obtained by using IN8 spectrometer. 
(f) Observed phonon dispersions of $\kappa$-Cl at $T=5$~K.
Thick lines in (f) are guide to the eyes.
}
\end{figure*}

Figure~\ref{fig1} shows contour map of phonon scattering intensity of 
$\kappa$-(BEDT-TTF)$_{2}$Cu[N(CN)$_{2}$]Cl at $T=5$~K.
Phonon spectra below 12~meV consist of several optical branches:
a low energy mode at $2\sim3$~meV, dispersive modes at $4\sim8$~meV
and zone-boundary modes at $>9$~meV.
Phonon signals are fitted to 
damped harmonic oscillator functions convoluted with instrumental resolution for constant-Q cuts, 
as shown in Figs.~\ref{fig2} (a)-(e). 
Obtained phonon dispersions are summarized in Fig.~\ref{fig2}(f).
At $\Gamma$-point, we observed four optical modes below 12~meV 
as shown in Fig.~\ref{fig2}(a).
The three modes at $E=6$, 8, and 11~meV at $\Gamma$-point show similar dispersions:
gradual decrease with increasing $q$.
On the other hand, the low-lying optical modes at 2~meV at the $\Gamma$-point
exhibit opposite $q$-dependence to the other modes 
and merge with the second lowest mode at $q=0.375$ (rlu).
Low-lying acoustic mode, presumably below 1.5~meV, was not observable
due to a large tail from incoherent elastic signal.

\begin{figure}
\includegraphics[keepaspectratio=true,width=7cm,clip]{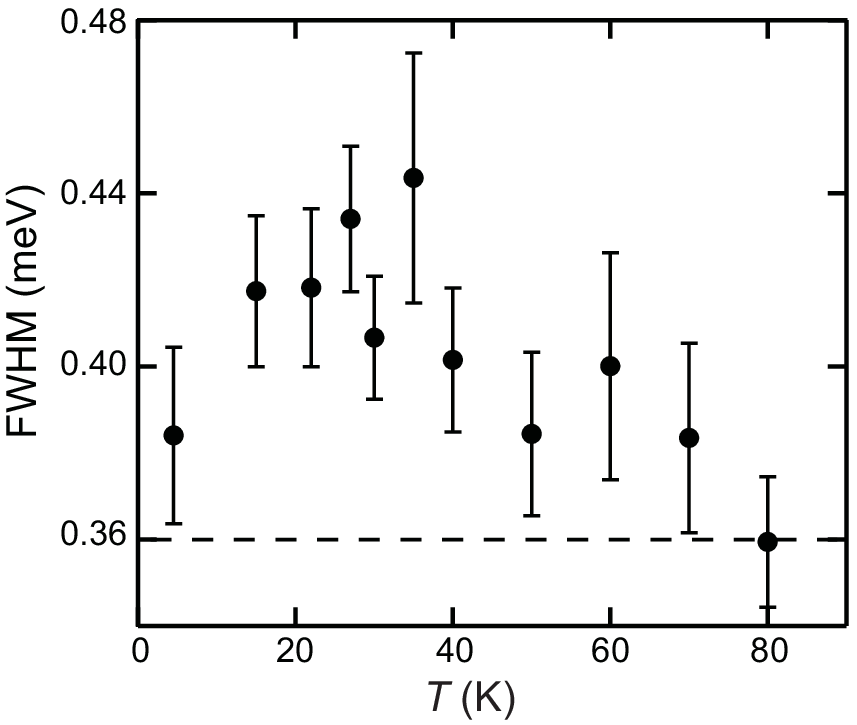}
\caption{\label{fig3} 
Temperature dependence of the energy width 
(Full-Width-at-Half-Maxium; FWHM) of the (603) Bragg peak.
The broken line shows the instrumental energy resolution.
Data in Fig.~\ref{fig3} were measured at PUMA spectrometer.
}
\end{figure}

Figure~\ref{fig3} shows the temperature dependence of the
energy width  (Full-Width-at-Half-Maxium; FWHM)  of the (603) peak.
The energy width  at $T=80$~K is almost resolution-limited 
$0.36\pm0.015$~meV.
At near $T_\mathrm{N}$,
the (603) peak is broadened in energy:
the energy width
increases to 0.44$\pm0.03$~meV at $T=35$~K 
then goes back to $0.38\pm0.02$~meV at $T=5$~K.
The broader peak than the instrumental energy resolution,
suggesting an unstable lattice for $T_\mathrm{N}<T<75$~K. 

\begin{figure}
\includegraphics[keepaspectratio=true,width=8cm,clip]{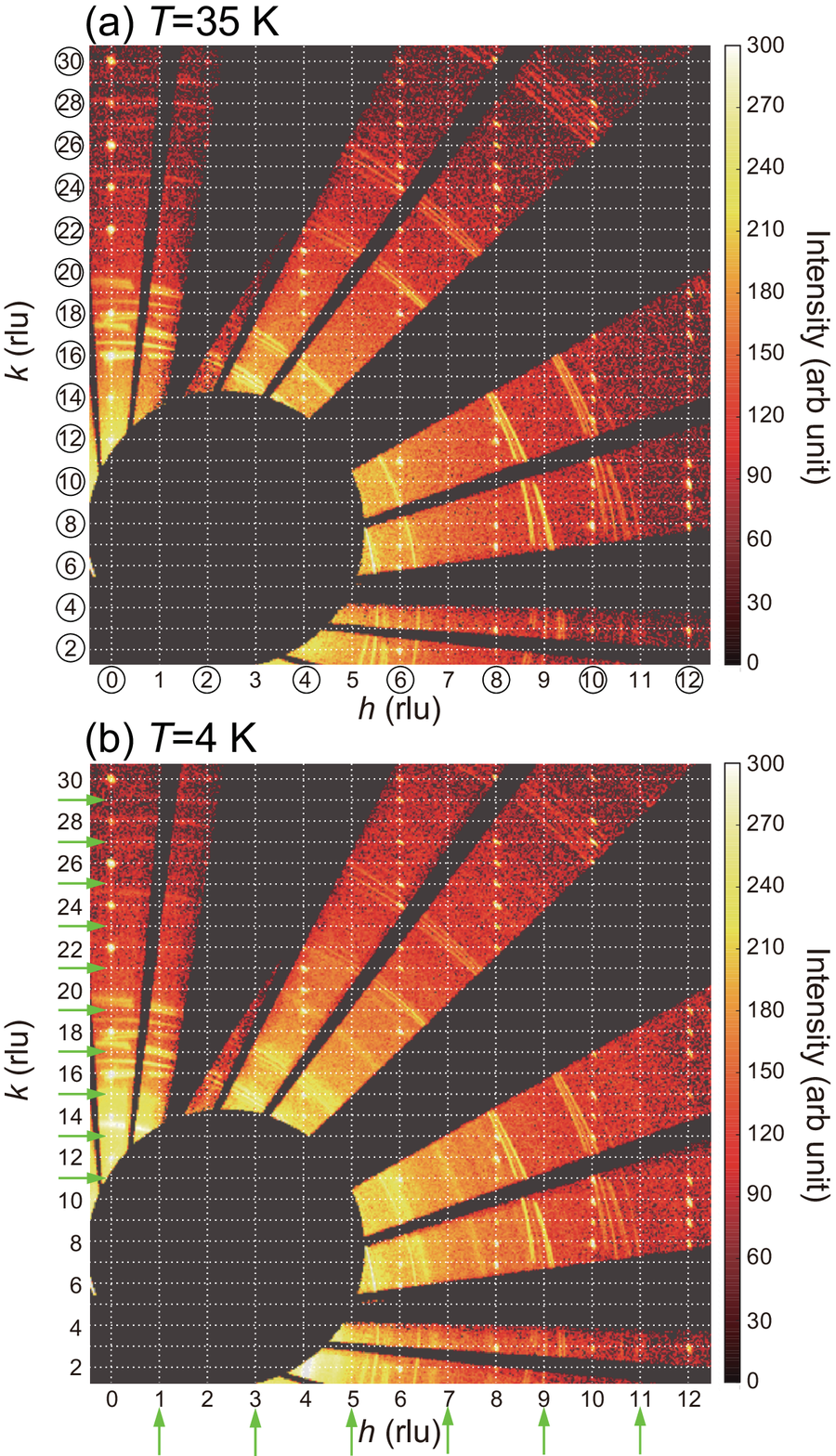}
\caption{\label{fig4}
Contour map of neutron diffraction intensity within the reciprocal space defined by the
orthogonal [1 0 0] and [0 1 0] axes of deuterated $\kappa$-Cl measured at (a) $T=35$~K
and (b) 4~K. The indexes with circle in (a) denote the reflection conditions: 
($hk0$) with $h=2n$ ($n$ is integer) 
and ($0k0$) with  $k=2n$ for the space group of Pnma.
The arrows in (b) indicate indexes of expected additional reflections 
($hk0$) with $h=2n+1$ and ($0k0$) with $k=2n+1$ for the acentric subgroups of Pnma.
}
\end{figure}

In order to check for the possibility of a change in the crystallographic symmetry at $T_\mathrm{N}$,
we performed detailed neutron diffraction measurements on a deuterated single crystal of $\kappa$-Cl
at $T=35$~K ($>T_\mathrm{N}$) and 4~K ($<T_\mathrm{N}$)
by using the Extreme Environment Single Crystal Neutron Diffractometer~\cite{SENJU}
SENJU installed at the Japan Proton Accelerator Research Complex
Materials and Life Science Experimental Facility.
Figure~\ref{fig4} show the intensity contour maps in the ($hk0$) scattering plane measured
at (a) $T=35$~K and (b) 4~K.
The Bragg peaks at $T=35$~K appear at ($hk0$) with $h=2n$ and ($0k0$) with $k=2n$,
where the indexes are enclosed within circles,
consistent with the space group Pnma reported in the literature~\cite{Williams90}.
Among the four acentric subgroups of Pnma:
P2$_{1}$2$_{1}$2$_{1}$, Pna2$_{1}$, Pnm2$_{1}$, and Pma2$_{1}$,
only Pma2$_{1}$ has the same reflection condition as Pnma.
The other acentric space groups give additional reflections ($hk0$) with $h=2n+1$ 
and ($0k0$) with $k=2n+1$, as indicated by arrows in Fig.~\ref{fig4}(b).
At $T=4$~K,
no additional superlattice peaks at these expected reflections were detected 
within an upper limit of $10^{-2}$ of the intensity of the main Bragg peaks. 
Moreover, the reflection conditions do not change from those at $T=35$~K.
Whereas a structural analysis assuming Pma2$_{1}$ for the 4~K data
yields strongly anomalous thermal factors, the Pnma model for the 4~K data converges with reasonable thermal factors.
Although we cannot rule out the existence of very weak superlattice peaks for $T<T_\mathrm{N}$,
possible structural changes due to inequivalent ET molecules within the dimer should be quite small.
The lack of any indications for a crystallographic symmetry lowering at $T_\mathrm{N} = T_{\rm FE}$ supports the picture
of electronic ferroelectricity, where instead of the lattice, the spin- or charge degrees of freedom are
the driving force of the phase transition in $\kappa$-Cl.